\documentclass[12pt, a4paper]{article}
\usepackage[height=25cm,width=16cm]{geometry}
\usepackage{feynmp}
\usepackage{amsmath}
\usepackage{ascmac}
\usepackage{dcolumn}
\usepackage{bm,here}
\usepackage[FIGTOPCAP]{subfigure}
\usepackage{comment}
\usepackage{ifpdf}
\usepackage{slashed}
\usepackage{colortbl}
\usepackage{color}
\usepackage{ulem}
\usepackage{comment}
\usepackage{braket}
\usepackage[mathscr]{eucal}
\usepackage[sort&compress, numbers, merge]{natbib}

\usepackage{cancel}

\ifpdf
\usepackage{graphicx}     
\usepackage[bookmarksopen,colorlinks=true,linkcolor=bblue,citecolor=ppink,urlcolor=ppink]{hyperref}
\else     
\usepackage[dvipdfmx]{graphicx}     
\usepackage[dvipdfmx,bookmarksopen, colorlinks=true, linkcolor=bblue, citecolor=ppink, urlcolor=ppink]{hyperref}
\fi
\usepackage{subcaption}
\usepackage{multicol}
\definecolor{red}{rgb}{1,0,0}
\definecolor{ppink}{rgb}{0.921545,0.440586,0.687243}
\definecolor{bblue}{rgb}{0.400000,0.400000,1.000000}
\usepackage[charter]{mathdesign}
\usepackage{soul}
\usepackage{wrapfig}

\begin{document}

\begin{titlepage}
\begin{center}

\vskip 1.5cm
{\Large \bf Féeton dark matter above the $e^-e^+$ threshold}

\vskip 2.0cm
{\large
Tatsuya Hayashi$^{(a)}$,
Shigeki Matsumoto$^{(a)}$,
Yuki Watanabe$^{(a)}$, \\ [.3em]
and
Tsutomu T. Yanagida$^{(a, b)}$
}

\vskip 1.5cm
$^{(a)}$ {\sl Kavli IPMU (WPI), UTIAS, University of Tokyo, Kashiwa, Japan} \\[.3em]
$^{(b)}$ {\sl Tsung-Dao Lee Institute, School of Physics and Astronomy, \\ Shanghai Jiao Tong University, China}\\[.1em]

\vskip 3.5cm
\begin{abstract}
\noindent
The new gauge boson introduced in the minimal extension of the standard model (SM) by gauging the U(1)$_{\rm B-L}$ symmetry plays the role of dark matter when the U(1)$_{\rm B-L}$ gauge coupling is highly suppressed. This dark matter, named the Féeton dark matter is known to be efficiently created in the early universe by inflationary fluctuations with minimal gravity coupling, hence the framework, the gauged U(1)$_{\rm B-L}$ extended SM + inflation, solves the four major problems of the SM; neutrino masses/mixings, dark matter, baryon asymmetry of the universe, and the initial condition of the universe (inflation). We comprehensively study the phenomenology of the dark matter when it is heavier than the $e^- e^+$ threshold, namely twice the electron mass, considering the threshold effect on the dark matter decay into $e^- e^+$. The viable parameter region is found only in the threshold region, while the branching fraction of the decay into $e^- e^+$ (i.e., the $e^- e^+$ signal) never vanishes even at the threshold due to the effect. As a result, the pure U(1)$_{\rm B-L}$ extension without the kinetic mixing between the U(1)$_{\rm B-L}$ and hyper-charge gauge bosons have already been excluded by the present observation of the 511\,keV photon from the galactic center. So, the Féeton dark matter requires a non-zero kinetic mixing to be a viable dark matter candidate and will be efficiently explored by future MeV-gamma ray telescopes thanks to the non-vanishing decay process into $e^- e^+$.
\end{abstract}

\end{center}
\end{titlepage}

\tableofcontents
\newpage
\setcounter{page}{1}

\section{Introduction}

The Standard Model (SM) of particle physics describes elementary particles and their interactions, and it has succeeded in giving explanations for almost all phenomena since its establishment\,\cite{ParticleDataGroup:2022pth}. However, there are problems that the SM cannot answer, which are now central challenges in modern physics. Some of the problems are particularly noteworthy in that the SM fails to explain certain experimental and observational results satisfactorily:
Precise observation of the Cosmic Microwave Background (CMB)\,\cite{Planck:2018jri} suggests that our universe seems to have undergone an exponential expansion in the very early epoch, known as the {\it inflation}\,\cite{Achucarro:2022qrl}. Furthermore, to generate the baryon asymmetry observed in the present universe from the symmetrically generated matter and antimatter due to the inflation, there must have been {\it baryogenesis} at some point in the history of our universe\,\cite{Bodeker:2020ghk}. In addition, the existence of the {\it neutrino masses and mixings}, as indicated by neutrino oscillation experiments\,\cite{Esteban:2020cvm}, and the {\it dark matter}, as suggested by many cosmological and astrophysical observations\,\cite{Cirelli:2024ssz}, are also significant problems requiring the extension of the SM.

Gauging the U(1)$_{\rm B-L}$ symmetry in the SM is known as the simplest solution to the problems of neutrino masses/mixings and baryogenesis mentioned above\,\cite{ParticleDataGroup:2022pth, Basso:2011hn}; three right-handed neutrinos are introduced to make the theory anomaly-free, which resolve the former problem via the seesaw mechanism\,\cite{Minkowski:1977sc, Yanagida:1979as, Yanagida:1979gs, Gell-Mann:1979vob}, while their decays in the early universe generate enough baryon asymmetry via the leptogenesis mechanism\,\cite{Fukugita:1986hr, Buchmuller:2005eh}. Moreover, recent studies have shown that the gauge boson introduced in the U(1)$_{\rm B-L}$ extension of the SM could serve as dark matter when the U(1)$_{\rm B-L}$ gauge coupling is highly suppressed. Importantly, enough amount of dark matter is created in the early universe during the inflation without contradicting any cosmological constraints, despite all interactions of the gauge boson being highly suppressed\,\cite{Graham_2016}. So, the model, the minimal U(1)$_{\rm B-L}$ extension of the SM with the inflation, is well-motivated, and its phenomenology is worth being studied in detail. The U(1)$_{\rm B-L}$ gauge boson playing the role of dark matter is called the {\it Féeton} dark matter.

The suppressed U(1)$_{\rm B-L}$ gauge coupling requires the Féeton dark matter to be lighter than ${\cal O}$(MeV). Then, the nature of the dark matter depends significantly on its mass; the dark matter annihilates mainly into neutrinos when it is lighter than the $e^-e^+$ threshold, i.e., twice the electron mass, while it also annihilates into an electron and a positron when it is heavier than the threshold, giving a strong signal in indirect dark matter detection. Moreover, the energy of the positrons from the dark matter decay is ${\cal O}$(MeV), so those efficiently capture ambient electrons in space, form positroniums, and contribute to the 511\,keV signal observed today\,\cite{Lin_2022,cheng2024feeton}. In this article, we consider the Féeton dark matter whose mass is greater than the $e^- e^+$ threshold and discuss its phenomenology in detail. We particularly focus on the following two effects that were not involved in past literature: the threshold (Sommerfeld) effect on the dark matter decay into $e^-e^+$ and that of the mixing between the kinetic terms of the U(1)$_{\rm B-L}$ and hyper-charge gauge bosons. We find that the viable parameter region of the dark matter is obtained in the threshold region due to constraints from astrophysical and cosmological observations. So, to quantitatively clarify the region, the decay width of the dark matter into $e^- e^+$ must be properly estimated, including the Sommerfeld effect: It turns out that the branching fraction of this decay never vanishes even at the threshold, $m_\mathfrak{f} = 2m_e$, with $m_\mathfrak{f}$ and $m_e$ being the masses of Féeton dark matter and electron.
As a result, the model without kinetic mixing is excluded; i.e., non-zero kinetic mixing is required to find the viable parameter region (to suppress the decay width).

This article is organized as follows: We introduce the minimal U(1)$_{\rm B-L}$ model in the next section\,(section\,\ref{sec: Feeton dark matter}) and discuss the nature of the Féeton dark matter. In section\,\ref{sec: Detection}, we consider the phenomenology of the Féeton dark matter, especially focusing on the constraint from astrophysical and cosmological observations based on a naive perturbative calculation of the decay width into $e^- e^+$. It then turns out that the viable parameter region can be found at the threshold regions. So, in section\,\ref{subsec: Decay with Sommerfeld}, we reevaluate the viable region, including the Sommerfeld effect on the decay. Finally, we summarize our study and briefly discuss a possible future direction in section\,\ref{sec: summary and future direction}. The meaning of the kinetic mixing between U(1)$_{\rm B-L}$ and U(1)$_Y$\.(hyper-charge) in the model and the derivation of the potential non-relativistic Lagrangian for the decay width calculation are also presented in appendices\,\ref{app: B-L+xY} and \ref{app: about pNR}. 

\section{Féeton dark matter}
\label{sec: Feeton dark matter}

The Féeton dark matter is nothing but the gauge boson associated with the U(1)$_{\rm B-L}$ gauge symmetry, where its stability is guaranteed by making the corresponding gauge coupling $g_{\rm B-L}$ small enough. The minimal extension of the standard model\,(SM) with this symmetry, known as the minimal U(1)$_{\rm B-L}$ model\,\cite{Jenkins:1987ue,Buchmuller:1991ce,Khalil_2008,Basso:2008iv}, is described by the following Lagrangian:
\begin{align}
    {\cal L}_{\rm B-L} &
    ={\cal L}_{\rm SM}
    -\frac{1}{4} V^{\mu\nu} V_{\mu\nu}
    +|D_\mu \Phi|^2
    +\sum_{i = 1}^3 \bar{N}_i\,i \gamma^\mu D_\mu N_i
    +V(\Phi,H)
    \nonumber \\
    &
    +g_{\rm B-L} V_\mu J_{\rm B-L}^\mu
    -\frac{\xi}{2}B_{\mu \nu} V^{\mu \nu}
    -\sum_{i, j = 1}^3
    \left[
        y_{ij}^{(\nu)} \bar{L}_i H^c N_j
        +\frac{1}{2} y_{ij}^{(N)} \bar{N}_i^c N_j \Phi
        + h.c.
    \right],
    \label{eq: lagrangian}
\end{align}
where ${\cal L}_{\rm SM}$ is the SM Lagrangian and $V_\mu$ is the U(1)$_{\rm B-L}$ gauge field with $V_{\mu\nu}$ being its field strength tensor. Meanwhile, $\Phi$ and $N_i$ are the new "Higgs" field breaking the U(1)$_{\rm B-L}$ symmetry spontaneously and the right-handed neutrinos that are introduced to make the model anomaly-free, with $D_\mu$ being their covariant derivatives. Here, U(1)$_{\rm B-L}$ charges of $\Phi$ and $N_i$ are $2$ and $-1$, respectively. The scalar potential among the new "Higgs" field $\Phi$ and the SM Higgs doublet field $H$ is denoted by $V(\Phi, H)$. The first term in the second line of the above Lagrangian is the U(1)$_{\rm B-L}$ gauge interactions of the SM fermions with the B$-$L current,
\begin{align}
    J_{\rm B-L}^\mu = \sum_f q_f \bar{f} \gamma^\mu f,
\end{align}
where $f = Q_i$, $U_i$, $D_i$, $L_i$, and $E_i$, and those are the quark doublet, up-type quark singlet, down-type quark singlet, lepton doublet, and charged lepton singlet, respectively, with the generation index being "$i$." The U(1)$_{\rm B-L}$ charge of the fermion $f$ is denoted by $q_f$, which is assigned to be $q_{Q_i} = q_{U_i} = q_{D_i} = 1/3$ and $q_{L_i} = q_{E_i} = -1$. The second term in the second line is the kinetic mixing between $V_\mu$ and the hyper-charge gauge boson $B_\mu$, with $B_{\mu \nu}$ being its field strength tensor.\footnote{
    The above minimal U(1)$_{\rm B-L}$ model with a non-zero kinetic mixing is indistinguishable from the U(1)$_{\rm B-L+xY}$ extension of the SM without the kinetic mixing between the new gauge boson and the SM hypercharge gauge boson when $\epsilon \ll 1$ and we choose the real number $x$ appropriately\,\cite{Lee:2016ief}. See appendix\,\ref{app: B-L+xY} for more details.}
Finally, terms in the bracket in the second line are Yukawa interactions concerning neutrinos $\nu_i \subset L_i$ and $N_i$, with their Yukawa couplings being $y_{ij}^{(\nu)}$ and $y_{ij}^{(N)}$.

The kinetic mixing term proportional to $\xi$ in the above Lagrangian makes $V_\mu$ mixed with the SM gauge fields $B_\mu$ and also $W^3_\mu$ via the electroweak symmetry breaking. Here, $W^3_\mu$ is the electrically neutral component of the SU(2)$_L$ gauge field multiplet. By taking the unitary gauge, the quadratic terms of the neutral fields $W^3_\mu$, $B_\mu$, and $V_\mu$ are obtained as follows:
\begin{align}
    &{\cal L}_{\rm B-L} \supset
    \frac{1}{2} (W^3_\mu, B_\mu, V_\mu)
    \left[ (\Box g^{\mu \nu} - \partial^\mu \partial^\nu)\,{\cal K} + {\cal M}\,g^{\mu \nu} \right]
    (W^3_\nu, B_\nu, V_\nu)^T,
    \nonumber \\
    & {\cal K} \equiv
    \begin{pmatrix} 1 & 0 & 0 \\ 0 & 1 & \xi \\ 0 & \xi & 1 \end{pmatrix},
    \qquad
    {\cal M} \equiv
    \begin{pmatrix} g^2 v_{\rm EW}^2/4 & -g g' v_{\rm EW}^2/4 & 0 \\ -g g' v_{\rm EW}^2/4 &  g^{\prime 2} v_{\rm EW}^2/4 & 0 \\ 0 & 0 & m_\mathfrak{f}^2 \end{pmatrix},
\end{align}
where $g$ and $g'$ are the SU(2)$_L$ and U(1)$_Y$ gauge couplings of the SM, $v_{\rm EW} \simeq$ 246\,GeV is the vacuum expectation value of $H$, and $m_\mathfrak{f} = 2 g_{\rm B-L} v_{\rm B-L}$ is the mass of the Féeton dark matter with $v_{\rm B-L}$ being the vacuum expectation value of $\Phi$. The redefinition of the fields and the diagonalization of the mass matrix give mass eigenstates with canonical kinetic terms,
\begin{align}
    &
    X {\cal K} X^T = {\bf 1},
    \qquad
    X {\cal M} X^T = {\rm diag}(m_Z^2, 0, m_\mathfrak{f}^2),
    \qquad
    (Z_\mu, A_\mu, \mathfrak{f}_\mu)^T = (X^{-1})^T (W^3_\mu, B_\mu, V_\mu)^T,
    \nonumber \\
    & \qquad\qquad
    X =
    \begin{pmatrix}
        \cos \theta_W & -\sin \theta_W & \displaystyle \xi \frac{\sin \theta_W m_Z^2}{m_Z^2 - m_\mathfrak{f}^2} \\
        \sin \theta_W & \cos \theta_W & 0 \\
        \displaystyle -\xi \frac{\cos \theta_W \sin \theta_W m_Z^2}{m_Z^2 - m_\mathfrak{f}^2} & \displaystyle \xi \frac{m_\mathfrak{f}^2 - \cos^2 \theta_W m_Z^2}{m_Z^2 - m_\mathfrak{f}^2} & 1
    \end{pmatrix}.
\end{align}
Here, $m_Z = (g^2 + g^{\prime 2})^{1/2}\,v_{\rm EW}/2$ is the mass of the $Z$ boson. We assumed that the mixing parameter is small ($\xi \ll 1$) and neglected terms smaller than ${\cal O}(\xi)$ to derive the above result. Assuming that the Féeton dark matter has a mass much smaller than the electroweak scale\,(i.e., $m_\mathfrak{f} \ll v_{\rm EW}$), it is found to have the following interactions with the SM particles:
\begin{align}
    {\cal L}_{\rm B-L}
    &\supset \sum_i
    \left[
        g_\nu \bar{\nu}_{L,i} \slashed{\mathfrak{f}} \nu_{L,i}
        + g_{e_L} \bar{e}_{L,i} \slashed{\mathfrak{f}} e_{L,i}
        + g_{e_R} \bar{e}_{R,i} \slashed{\mathfrak{f}} e_{R,i}
    \right] + \cdots, \\
    &g_\nu \simeq
    -g_{\rm B-L}
    -g' \xi\,\frac{m_\mathfrak{f}^2}{2 (m_Z^2 - m_{\mathfrak{f}}^2)} \simeq -g_{\rm B-L} \\
    &g_{e_L} \simeq
    -g_{\rm B-L}
    +g' \xi\,\frac{c_W^2 m_Z^2 - m_{\mathfrak{f}}^2}{m_Z^2 - m_{\mathfrak{f}}^2}
    +g' \xi\,\frac{m_{\mathfrak{f}}^2}{2 (m_Z^2 - m_{\mathfrak{f}}^2)}
    \simeq -g_{\rm B-L} +  \epsilon\,e \\
    &g_{e_R} \simeq
    -g_{\rm B-L}
    +g' \xi\,\frac{c_W^2 m_Z^2 - m_{\mathfrak{f}}^2}{m_Z^2 - m_{\mathfrak{f}}^2}
    \simeq -g_{\rm B-L} +  \epsilon\,e,
\end{align}
with $c_W = \cos\theta_W$, $e=c_W g'$ and $\epsilon = c_W \xi$. It is seen that the interactions between the Féeton dark matter and charged leptons become vector-like whenever $m_\mathfrak{f} \ll m_Z$. It is also true for interactions between the Féeton dark matter and quarks. However, we do not explicitly write the interactions and those with other heavy fields (e.g., weak gauge bosons, SM and U(1)$_{\rm B-L}$ Higgs bosons) because they are irrelevant to the following discussion.

When the mass of the Féeton dark matter is light enough, say less than $\sim$100\,MeV, it decays mainly into a pair of neutrinos and a pair of electrons. The partial decay widths of these decay modes are obtained at the leading order of the perturbation as follows:
\begin{align}
    &\Gamma(\mathfrak{f} \to \nu_i \bar{\nu}_i) =
    \frac{g_{\rm B-L}^2}{24\pi} m_\mathfrak{f},
    \label{eq: f to nunu at LO} \\
    &\Gamma(\mathfrak{f} \to e^- e^+) =
    \frac{(g_{\rm B-L}-\epsilon\,e)^2}{12\pi} m_{\rm f}
    \left( 1 + \frac{2 m_e^2}{m_\mathfrak{f}^2} \right)
    \left( 1 - \frac{4 m_e^2}{m_\mathfrak{f}^2} \right)^{1/2}
    \label{eq: decay width electron}
\end{align}
with $m_e$ being the electron mass. The partial decay width\,(\ref{eq: f to nunu at LO}) is the one for each neutrino flavor, so the decay width into a pair of neutrinos in all the flavors is three times larger than that in eq.\,(\ref{eq: f to nunu at LO}). So, the total decay width of the Féeton dark matter is $\Gamma_{\rm tot} \simeq [3\,g_{\rm B-L}^2 + 2(g_{\rm B-L} - \epsilon e)^2]\,m_\mathfrak{f}/(24 \pi)$ assuming $m_\mathfrak{f} \gg m_e$, where its inverse, i.e., the lifetime of the Féeton dark matter, must be longer enough than the age of the universe, $\tau_U \simeq 4.3 \times 10^{17}$\,s:
\begin{align}
    \Gamma_{\rm tot}^{-1} \gg \tau_{\rm U}
    \qquad \Rightarrow \qquad
    g_{\rm B-L}^2\,m_\mathfrak{f} \ll 1.2 \times 10^{-38}\,{\rm MeV}.
    \label{eq: lifetime}
\end{align}
On the other hand, the Féeton dark matter mass is $m_\mathfrak{f} = 2 g_{\rm B-L} v_{\rm B-L}$ as addressed above, where the vacuum expectation value $v_{\rm B-L}$ must be smaller than the Planck scale, $m_{\rm pl} \simeq 2.4 \times 10^{18}$\,GeV, to justify that the physics of the dark matter is described within the framework of the quantum field theory. This condition gives an inequality on $g_{\rm B-L}$ and $m_\mathfrak{f}$ as
\begin{align}
    v_{\rm B-L} \lesssim m_{\rm pl}
    \qquad \Rightarrow \qquad
    g_{\rm B-L}^{-1}\,m_\mathfrak{f} \lesssim 4.8 \times 10^{21}\,{\rm MeV}.
    \label{eq: vev}
\end{align}
The above two inequalities in eqs.\,(\ref{eq: lifetime}) and (\ref{eq: vev}) give us the upper limit on $m_\mathfrak{f}$ as $m_\mathfrak{f} \ll {\cal O}(100)$\,MeV, and the upper limit on $g_{\rm B-L}$ as $g_{\rm B-L} \ll {\cal O}(10^{-19})$, assuming $m_\mathfrak{f} > 2 m_e$.\footnote{
    The inequalities also give a lower limit on $v_{\rm B-L}$ as $v_{\rm B-L} \gtrsim 10^{16}$\,GeV. Meanwhile, the inflationary fluctuation creates the Féeton dark matter abundance as $\Omega_\mathfrak{f}\,h^2 \simeq 0.05\,(m_\mathfrak{f}/1\,{\rm MeV})^{1/2} (H_I/10^{11}\,{\rm GeV})^2$ with $H_I$ being the inflationary scale\,\cite{Graham_2016}. Since the dark matter abundance observed today is $\Omega_{\rm DM}\,h^2 \simeq 0.12$, the scale $H_I$ must be around $10^{11}$\,GeV, which is smaller than the lower limit on $v_{\rm B-L}$, validating this dark matter creation. Unfortunately, the inflationary tensor mode is difficult to observe, as the inflationary scale $H_I$ is small.}
So, the MeV scale emerges inherently as the maximally allowed mass of the Féeton dark matter.

\section{Detecting the Féeton dark matter}
\label{sec: Detection}

Next, we consider the detection of the Féeton dark matter. As seen in the previous section, the dark matter has very suppressed interactions with the SM particles. Hence, the usual search strategies for a particle dark matter candidate utilizing collider and underground experiments do not work. So, the detection must rely on indirect detection utilizing astrophysical and cosmological observations; we observe products in the present universe and impacts on the early universe cosmology caused by the decay of the dark matter.

\begin{figure}[t]
    \centering
    \includegraphics[keepaspectratio, scale=0.5]{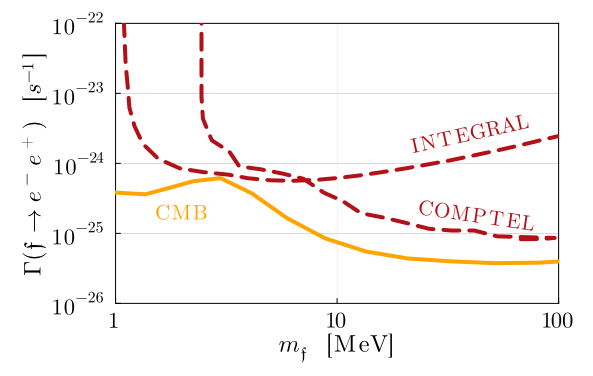}
    \caption{\small \sl Upper bounds on the partial decay width of the Féeton dark matter into a pair of electrons, $\Gamma(\mathfrak{f} \to e^-e^+)$, at 95\,\% C.L., which are obtained from the precise CMB observation (solid orange line), as well as present-day diffuse photon searches (dotted brown lines). See text for more details.}
    \label{fig: constraint on ee}
\end{figure}

Let us first consider the impact of the decay on the early universe cosmology. Since the Féeton dark matter decays into a pair of electrons at the tree level, it may affect the physics of the thermal bath composed of the SM particles (electron, positron, and photon) in the early universe, i.e., the thermal history of the early universe such as the recombination\,\cite{Slatyer_2017} and the reionization\,\cite{Liu_2016}. However, cosmological observation prefers history without such a new physics contribution and constrains the partial decay width of the dark matter into a pair of electrons. Among various constraints, the most robust and stringent constraint is obtained from the physics of the recombination\,\cite{Slatyer_2017}, and the precise CMB observation gives an upper bound on the partial decay width as shown in Fig.\,\ref{fig: constraint on ee}. It is seen in the figure that the decay width, $\Gamma(\mathfrak{f} \to e^-e^+)$, must be smaller than $\sim 10^{-25}$\,s$^{-1}$ at 95\,\% C.L. Combining the above upper bound on the partial decay width with the limit shown in eq.\,(\ref{eq: vev}) gives an upper bound on the branching fraction of the dark matter decay into a pair of electrons. The equation\,(\ref{eq: vev}) gives a lower bound on the total decay width as follows:
\begin{align}
    \Gamma_{\rm tot}
    \geq \frac{g_{\rm B-L}^2}{8\pi} m_{\mathfrak{f}}
    = \frac{1}{32\pi v_{\rm B-L}^2} m_{\mathfrak{f}}^3
    \geq \frac{1}{32\pi m_{\rm pl}^2} m_{\mathfrak{f}}^3
    > \frac{1}{4 \pi m_{\rm pl}^2} m_e^3
    \simeq 3 \times 10^{-24}\,{\rm s}^{-1},
\end{align}
where we use the inequality, $m_{\mathfrak{f}} > 2m_e$, as we focused on such a mass region in this article. So, the branching fraction of the dark matter into $e^- e^+$ has the following upper bound:
\begin{align}
    {\rm Br}(\mathfrak{f} \to e^- e^+)
    \equiv \frac{\Gamma(\mathfrak{f} \to e^- e^+)}{\Gamma_{\rm tot}}
    \lesssim 3 \times 10^{-2}.
\end{align}
The branching fraction into a pair of electrons is at most ${\cal O}(10)$\,\% level, so it cannot be a dominant decay process; the Féeton dark matter decays mainly into a pair of neutrinos. The decay of the Féeton dark matter into a pair of neutrinos may also affect the early universe cosmology. When a substantial amount of dark matter decays in the early universe, it changes the energy budget of the universe and hence the Hubble parameter, modifying the early universe cosmology. The precise CMB observation, as well as baryonic acoustic oscillation measurements, do not prefer the cosmology with such a new physics contribution; those give an upper bound on the total decay width of the dark matter as $\Gamma_{\rm tot} < 1.3 \times 10^{-19}$\,s at 95\,\% C.L., assuming the dark matter decays dominantly into a pair of neutrinos\,\cite{Alvi_2022}.

\begin{figure}[t]
    \begin{center}
    \subfigure[\qquad\quad$|g_{\rm B-L} - \epsilon e|/g_{\rm B-L}=1$]{
    \includegraphics[width=.482\columnwidth]{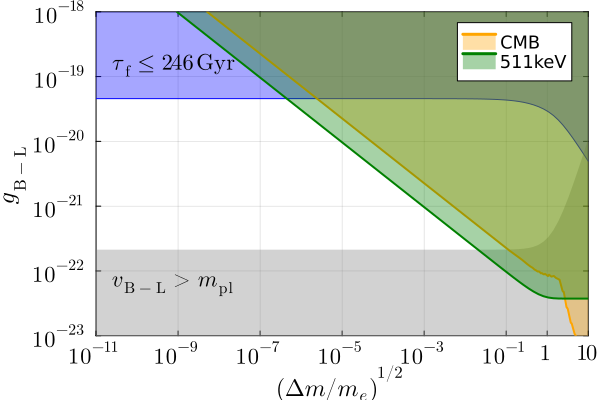}
    }
    \subfigure[\qquad\quad$|g_{\rm B-L} - \epsilon e|/g_{\rm B-L}=10^{-1}$]{
    \includegraphics[width=.482\columnwidth]{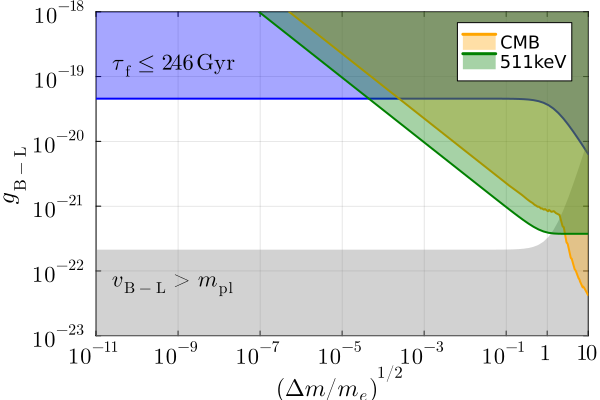}
    }
    \subfigure[\qquad\quad$|g_{\rm B-L} - \epsilon e|/g_{\rm B-L}=10^{-2}$]{
    \includegraphics[width=.482\columnwidth]{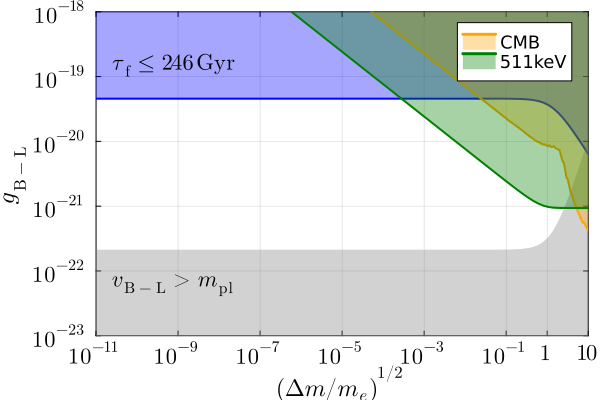}
    }
    \subfigure[\qquad\quad$(\Delta m/m_{e})^{1/2}=10^{-3}$]{
    \includegraphics[width=.482\columnwidth]{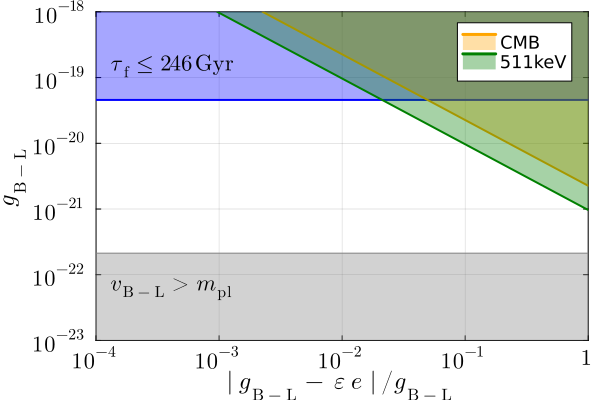}
    }
    \caption{\small \sl 
    {\bf Top and bottom-left panels}: The model parameter region consistent with the limits on $\Gamma(\mathfrak{f} \to e^-e^+)$ from the CMB and the 511\,keV observations (orange- and green-shaded regions), $\Gamma_{\rm tot}$ (blue-shaded region), and $v_{\rm B-L}$ (gray-shaded region), with several choices of $|g_{\rm B-L} - \epsilon e|/g_{\rm B-L}$. Here, $\Delta m \equiv m_{\mathfrak{f}} - 2m_e$. {\bf Bottom-right panel:} The region is depicted with $\Delta m/m_e$ being fixed as $10^{-3}$.}
    \label{fig:result_without_Sommerfeld}
    \end{center}
\end{figure}

The model parameter region that is consistent with the limits on $\Gamma(\mathfrak{f} \to e^-e^+)$, $\Gamma_{\rm tot}$, and eq.\,(\ref{eq: vev}) is shown in the top and bottom-left panels of Fig.\,\ref{fig:result_without_Sommerfeld}, with several choices of $|g_{\rm B-L} - \epsilon e|/g_{\rm B-L}$. Here, $\Delta m \equiv m_{\mathfrak{f}} - 2m_e$. In the bottom-right panel, the region is depicted with $\Delta m/m_e$ being fixed as $10^{-3}$. It is seen in the figure that the limits on $\Gamma_{\rm tot}$ and eq.\,(\ref{eq: vev}) require the mass of the Féeton dark matter to be below 10\,MeV. Furthermore, the limit on $\Gamma(\mathfrak{f} \to e^-e^+)$ from the CMB observation gives a more stringent limit, requiring the mass to be in the threshold region ($\Delta m/m_e \lesssim 1$\,MeV), depending on the $|g_{\rm B-L} - \epsilon e|/g_{\rm B-L}$ value.\footnote{
    The weak gravity conjecture allows $g_{\rm B-L} = {\cal O}(10^{-22}$--$10^{-19})$ thanks to the existence of neutrinos\,\cite{Harlow:2022ich, Benakli:2020vng}.}

Next, we consider the detection of the Féeton dark matter by observing its decay products in the present universe. Signals, i.e., the fluxes of the decay products, comprise three contributions in general, i.e., the primary, secondary, and tertiary contributions, as follows:
\begin{align}
     \frac{d^2F_i}{dE d\Omega} =
     \left. \frac{d^2F_i}{dE d\Omega} \right|_{\rm Prim} +
     \left. \frac{d^2F_i}{dE d\Omega} \right|_{\rm Sec} +
     \left. \frac{d^2F_i}{dE d\Omega} \right|_{\rm Tert},
\end{align}
with $E$ and $\Omega$ being the energy of the products and the direction from which the signal comes, while the index `i' represents the spices of the products, such as the electron/positron, photon, and neutrino. The {\bf primary contribution} is directly from the decay. In the case of the Féeton dark matter, it gives the signal fluxes of electrons/positrons, neutrinos, and photons. First, the electron/positron flux is difficult to detect, as the energies of the electrons/positrons from the decay are too small to penetrate our heliosphere. The only possible exception would be the detection by Voyager\,1 located outside the heliosphere; however, even the Voyager\,1 cannot detect the flux because its sensitivity is not good enough for $\lesssim {\cal O}(10)$\,MeV energy electrons/positrons\,\cite{Boudaud_2017}. Second, the neutrino flux is also difficult to detect, though the dark matter decay dominantly produces neutrinos; the expected neutrino flux is much below the current detection sensitivity\,\cite{cheng2024feeton}. Finally, photons are also produced from the decay via the final state radiation process with the differential decay width of
{\small
\begin{align}
    &
    \frac{d\Gamma(\mathfrak{f} \to e^- e^+ \gamma)}{dE} = 
    \frac{2 \alpha}{\pi m_\mathfrak{f}} \Gamma(\mathfrak{f} \to e^- e^+)
    \times {\rm FSR}(2E/m_\mathfrak{f}, m_e/m_\mathfrak{f}),
    \label{eq: eeg width}
    \\
    &
    {\rm FSR}(x, \mu) =
    \frac{1 + (1 - x)^2 - 4\mu^2 (x + 2 \mu^2)}{x\,\sqrt{1 - 4\mu^2}\,(1+2\mu^2)}
    \log\left[\frac{1 + v_\mu(x)}{1 - v_\mu(x)}\right]
    -\frac{1 + (1 - x)^2 + 4\mu^2 (1 - x)}{x\,\sqrt{1 - 4\mu^2}\,(1 + 2\mu^2)}
    v_\mu(x),
    \nonumber
\end{align}
}\noindent
with $v_\mu(x) \equiv [1 - 4 \mu^2/(1 - x)]^{1/2}$ and $\alpha$ being the fine structure constant\,\cite{Coogan_2020}. The partial width of this decay is obtained by integrating the above differential width over an appropriate period of the energy $E$.\footnote{
    Because of the infrared divergence (i.e., the divergence at $E \to 0$) in eq.\,(\ref{eq: eeg width}), which is regularized by an interference term between the leading and the next-leading order diagrams of the $\mathfrak{f} \to e^- e^+$ process, the partial decay width of $\Gamma(\mathfrak{f} \to e^- e^+ \gamma)$ is computed as that with a photon having e.g., $E \geq 10^{-3} m_e$. It means that $\Gamma(\mathfrak{f} \to e^- e^+ \gamma)$ with a photon having $E \leq 10^{-3} m_e$ is treated as a radiative correction to $\Gamma(\mathfrak{f} \to e^- e^+)$.}
However, the signal flux of the photon is not strong, especially at the threshold region $m_\mathfrak{f} \sim 2m_e$. The search for the photon signal at the INTEGRAL and COMPTEL observations put an upper bound on the partial decay width into a pair of electrons, $\Gamma(\mathfrak{f} \to e^- e^+)$, utilizing the equation\,(\ref{eq: eeg width})\,\cite{Essig_2013}; however, it is comparable or weaker than that obtained by the cosmological observation discussed above, as seen in Fig.\,\ref{fig: constraint on ee}.

The {\bf secondary contribution} is from the interaction of the decay products with the interstellar medium. In the Féeton dark matter case, it gives the signal flux of photons caused by the interaction between the electrons/positrons from the decay and the interstellar radiation (CMB, infrared radiation, and starlight) via the inverse Compton scattering process and between the electrons/positrons and the interstellar matter (hydrogen, helium, etc.) via the Bremsstrahlung process, etc. However, the strength of the signal is again not strong because the energies of injected electrons/positrons are less than ${\cal O}(10)$\,MeV and resultant secondary photons are too low in energy; the constraint on $\Gamma(\mathfrak{f} \to e^-e^+)$ obtained from the search for such secondary photons is weaker than those obtained from the search for the primary photons and from the cosmological observation discussed above.\,\cite{Cirelli:2020bpc, Cirelli:2023tnx}.\footnote{
    The constraint from the secondary contribution to the photon signal and that from the primary contribution to the electron/positron signal discussed in the previous paragraph depend strongly on how low-energy electrons/positrons propagate in our galaxy. Although such a preparation is not yet well known, it could boost the signals, especially when the so-called reacceleration of the electrons/positrons is efficient during the propagation\,\cite{luque2024importance}. We do not consider such a boost in this article to make the constraints conservative.}

The {\bf tertiary contribution} consists of those not included in the primary and secondary. Unlike the previous two contributions, in the case of Féeton dark matter, we expect a significant and characteristic tertiary contribution to the signal flux of photons: The decay of the dark matter produces positrons with energies less than 10\,MeV, many of those will form positroniums by capturing ambient electrons, and the positroniums decay into two photons with $\sim 25$\,\% branching fraction (i.e., the decay of para-positroniums). So, the dark matter decay eventually gives a monochromatic photon signal with the energy of $E_\gamma \simeq m_e \simeq 511$\,keV. On the observational side, the 511\,keV photon flux has already been detected from both galactic bulge and disk regions. According to the analysis of the INTEGRAL SPI data, the amount of the 511keV photon flux from the bulge region (a round area centered at the galactic center with a radius of $\sim 10.3^\circ$) is estimated to be $F_{511}^{\rm Obs} \simeq (4.8$--$9.6) \times 10^{-4}$\,ph\,cm$^{-2}$s$^{-1}$\,\cite{Siegert_2016}.\footnote{
    It turns out that the 511\,keV flux from the bulge region is fitted well by the two Gaussian components centered (almost) at the galactic center; one has an FWHM of 20.55$^\circ$, and another has an FWHM of 5.75$^\circ$\,\cite{Siegert_2016}. The relative contributions from the two components are not given, so we estimate the total flux from the 10.3$^\circ$ region assuming that the former\,(latter) dominates, which gives $F_{511}^{\rm Obs} \simeq 4.8\,(9.6) \times 10^{-4}$\,ph\,cm$^{-2}$s$^{-1}$.}
Moreover, the continuum photon flux from ortho-positroniums has also been detected: The ratio of the 511\,keV line flux to that of the ortho-positronium continuum gives the information of the positronium fraction $f_{\rm Ps}$, the fraction of $e^- e^+$ annihilation for the 511\,keV flux that occurs via the formation of a positronium atom. According to the observation, it turns out that the fraction is consistent with 100\,\% within several percent error, so the observed 511\,keV flux is mainly from the decay of the para-positroniums.

However, on the theoretical side, the origin of the observed 511\,keV flux from the galactic bulge region, i.e., the origin of the positron production in the region, is not well known; in addition to the dark matter decay, it could also be from microquasars, supernovae, and massive stars, though how much contributions those give is uncertain\,\cite{dissertation, Siegert_2023}. Furthermore, the contribution to the 511\,keV flux from the dark matter decay itself is very uncertain. First, it depends on how the dark matter is distributed in our galaxy, i.e., the dark matter profile in the bulge region: the dark matter contribution is proportional to the so-called $D$-factor,
\begin{align}
    D \equiv
    \int_{\rm Bulge} d\Omega \int_{\rm l.o.s} d\ell\,
    \rho_{\rm DM}(\ell, \Omega)
    = 2\pi \int_0^{10.3^\circ} \sin \theta\,d\theta \int_{\rm l.o.s} d\ell\,\rho_{\rm DM}(\ell, \Omega(\theta)),
\end{align}
with $\rho_{\rm DM}$ being the mass density of the dark matter towards the direction $\Omega$. Assuming various profiles such as NFW, Einasto, Burkert, etc., the $D$-factor takes a value of (3.4--11)$\times 10^{21}$\,GeV/cm$^2$, indicating ${\cal O}(100)$\,\% uncertainty exists\,\cite{Marco_Cirelli_2011}. Moreover, the contribution also depends on how the low-energy positrons propagate and how those capture ambient electrons to form positroniums: Positrons injected into our galaxy lose their energies during their propagation by collisions with electrons, interactions with plasma waves, excitation, ionization of atoms/molecules, and those are eventually thermalized to the interstellar medium composed of H\,(hydrogen atom), He\,(helium atom), H2\,(hydrogen molecule), grains, as well as ionized particles such as free electrons. Then, the positrons capture electrons during the propagation and after the thermalization via various charge exchange processes (e.g., $e^+ + {\rm H} \to {\rm Ps} + p$ with Ps being positronium) and radiative recombination with free electron\,\cite{Guessoum_2005}. Since both the propagation of low-energy positrons and the distribution of the medium inside the bulge region are not well known, it is difficult to predict the morphology of the 511\,keV flux from the dark matter decay within reasonable accuracy.

So, we focus on the total 511\,keV flux from the bulge region to constrain the Féeton dark matter parameters. Even in such a case, several uncertainties are associated with estimating the flux from the dark matter decay. First, as discussed above, the dark matter distribution is uncertain; we adopt the $D$-factor of $3.4 \times 10^{21}$\,GeV/cm$^2$ to make the constraint conservative. Next, some positrons injected into our galaxy annihilate with ambient electrons before forming positronium, i.e., the in-flight annihilation. Fortunately, the influence of the annihilation is negligibly small compared to others when the energy of the injected positrons is small enough\,\cite{Martin_2012}. Finally, all the positrons do not necessarily annihilate into photons inside the bulge region, i.e., some positrons may leave the region without experiencing annihilation.\footnote{
    The positrons escaping from the bulge region may capture electrons, form positroniums, and decay into photons outside the region. Unfortunately, those contribute to the (almost) isotropic diffused photon flux, which is difficult to detect at an X-ray telescope using a coded aperture mask such as INTEGRAL/SPI.}
On the other hand, evaluating how many positrons could be away from the bulge region is not easy; according to Ref.\,\cite{dissertation, Siegert_2023}, at most 80\,\% (94\,\%) of the positrons could be away from the bulge (disc) region at 2$\sigma$ level, assuming all positrons are produced from astrophysical activities. Because this estimate is from the comparison between the observed $e^+$ annihilation rate and the possible production rate of positrons in each region, it cannot be directly applied to the case with the dark matter decay. However, since the contribution of the dark matter decay to the 511\,keV flux from the disk region is sub-leading, the above estimate, i.e., at most 94\,\% of the positrons could be away from the region, can still be used. Moreover, assuming that the escaping rate from the bulge region is lower than that from the disc region even if the dark matter decay dominantly contributes to the positron production in the bulge region,\footnote{
    This statement is validated because the kinetic energy of the positrons produced by the dark matter decay is comparable to or smaller than the positrons from the astrophysical activities, as deduced from Fig.\,\ref{fig:result_without_Sommerfeld}.}
we can constrain the total 511 keV flux from the bulge region as
\begin{align}
    F_{511}^{\rm DM} < \frac{F_{511}^{\rm Obs}}{1 - 0.94},
    \label{eq: 511 limit}
\end{align}
where $F_{511}^{\rm Obs} \simeq 9.6 \times 10^{-4}$\,ph\,cm$^{-2}$s$^{-1}$ is a conserved estimate of the observed 511\,keV flux from the bulge region (as we discussed above), while $F_{511}^{\rm DM}$ is a conserved estimate of the theoretical prediction on the flux from the dark matter decay, which is obtained as
\begin{align}
    F_{511}^{\rm DM} =
    2 \times \frac{1}{4} \times f_{Ps}
    \frac{\Gamma( \mathfrak{f} \to e^- e^+)}{4\pi m_\mathfrak{f}} D,
    \label{eq: DM 511}
\end{align}
where factors 2 and 1/4 in the formula are from the number of photons produced in each positronium decay and ${\rm Ps}$ decay branching fraction into two photons, respectively.

The model parameter region excluded by the limit\,(\ref{eq: 511 limit}) is shown in Fig.\,\ref{fig:result_without_Sommerfeld} as a green-shaded region. It is found that the limit from the 511\,keV observation is stronger than that from the CMB observation, though we adopt a very conservative setup; it restricts the mass of the Féeton dark matter further to the threshold region. However, the decay width obtained by a perturbative calculation, i.e., that in eq.\,(\ref{eq: decay width electron}), is not reliable anymore in such a deep threshold region, $\Delta m/m_e \lesssim 0.1$\,MeV, due to the threshold singularity. So, we reevaluate the limits in the next section, including the effect of the singularity.

\section{Féeton dark matter decay at the \texorpdfstring{$e^- e^+$}{TEXT} threshold}
  
The perturbative calculation of a decay width at a threshold region is not validated when a long-range force exists among final state particles, known as the threshold singularity. The decay of the Féeton dark matter into an electrons pair in the threshold region is exactly the case: The width shown in eq.(\ref{eq: decay width electron}), which we adopted to evaluate the constraints in the previous section, is indeed invalid, as the correction from $n$-loop diagrams to the decay amplitude with a ${\cal O}(e\alpha^n)$ contribution is associated with an additional kinematical factor $(m_e/\Delta m_\mathfrak{f})^n$ from infrared regions of loop momenta. The factor enhances the amplitude at each order of the perturbation, and the naive perturbative computation of the decay amplitude breaks in the deep threshold region. In this section, we consider the decay of the Féeton dark matter into a pair of electrons at the threshold regions, including the threshold (Sommerfeld) effect, and reevaluate the constraints discussed in the previous section.

\subsection{Decay width \texorpdfstring{$\Gamma(\mathfrak{f} \to e^- e^+)$}{TEXT}}
\label{subsec: Decay with Sommerfeld}

In the final state, the electromagnetic interaction (exchanging photons) between an electron and a positron causes a long-range force at the threshold region, leading to the threshold singularity. We evaluate the Sommerfeld effect on the decay width using the potential non-relativistic (NR) Lagrangian method\,\cite{Pineda_1998, Brambilla_2000}, which is the effective field theory obtained by integrating hard ($\ell^0 \sim |\vec{\ell}| \sim m_e$) and soft ($\ell^0 \sim |\vec{\ell}| \sim \beta m_e$) parts, while keeping the potential ($\ell^0 \sim \beta^2 m_e$, $|\vec{\ell}| \sim \beta m_e$) and the ultra-soft ($\ell^0 \sim |\vec{\ell}| \sim \beta^2 m_e$) parts, of the electron and the photon fields out from the original Lagrangian, with $\beta$ being the electron velocity in the final state. At the leading order of $\beta$ and $\alpha$, the potential NR Lagrangian is obtained as
\begin{align}
    \mathcal{L}^{({\rm pNR})}_\mathfrak{f}
    &=
    -\frac{1}{2} \vec{\mathfrak{f}}\,(\Box +m_\mathfrak{f}^2)\,\vec{\mathfrak{f}}
    + \int d^3r\,\vec{\Phi}^\dagger(\vec{r},x) \left(i\partial_{x^0} + \frac{\nabla_x^2}{4m_e} + \frac{\nabla_r^2}{m_e} + \frac{\alpha}{r} \right)  \vec{\Phi}(\vec{r},x)
    \nonumber \\
    &-
    \sqrt{2} (g_{\rm B-L} - \epsilon e)\,
    \vec{\mathfrak{f}} \cdot
    \left[ e^{2i m_e x^0} \vec{\Phi}^\dagger (\vec{0}, x) + e^{-2i m_e x^0} \vec{\Phi}(\vec{0},x) \right],
    \label{eq: pNR Lagrangian}
\end{align}
where $\vec{\mathfrak{f}}$ is the spatial components of the Féeton dark matter field, while $\vec{\Phi}(\vec{r},x)$ and $\vec{\Phi}^\dagger(\vec{r},x)$ are the fields annihilating and creating the non-relativistic $e^- e^+$ two-body state with the total spin of one (i.e., the spin-triplet state), respectively. Here, $\vec{r}$ and $x$ are the relative and the center-of-mass coordinates of the $e^- e^+$ two-body system. It is seen that the Coulomb potential, $-\alpha/r$, exists in the Lagrangian.\footnote{
    The potential NR Lagrangian method can be applied only at the threshold region, i.e., in other regions away from the threshold, the use of the method is not validated. On the other hand, the Sommerfeld effect is suppressed in the regions, and the usual perturbative calculation can be used to compute the width.}
There are several ways to derive the above potential Lagrangian. Please see appendix\,\ref{app: about pNR}, where one of them is given in some detail.

It is convenient to expand the two-body field $\Phi(\vec{r},x)$ by the solutions of the Schr\"{o}dinger equation (i.e., the equation of motion for the two-body field describing the relative motion between an electron and a positron) to calculate the decay width of the Féeton dark matter including the Sommerfeld effect. Then, we expand the $\Phi(\vec{r},x)$ field as follows:
\begin{align}
    \vec{\Phi}(\vec{r},x) =
    \sum_{\ell,\,m} \int_0^\infty \frac{dk}{2\pi}\,\vec{C}_{k \ell m}(x)\,\psi_{k \ell m}(\vec{r})
    + \cdots,
    \label{eq: expansion}
\end{align}
where $\psi_{k \ell m}(\vec{r})$ is the solution describing a continuum state of the two-body system with the wave number $k = (m_e E)^{1/2}$, the azimuthal and magnetic quantum numbers $\ell$ and $m$, respectively, with $E$ being the internal (kinetic) energy of the state. On the other hand, the coefficient $\vec{C}_{k \ell m}(x)$ is the field annihilating the continuum state. The solution $\psi_{k \ell m}(\vec{r})$ is required to satisfy the specific normalization condition, $\int d^3r\,\psi^\dagger_{k' \ell' m'}(x)\,\psi_{k \ell m}(x) = (2\pi)\,\delta(k - k')\,\delta_{\ell \ell'}\,\delta_{m m'}$, to make the kinetic term of the $\vec{C}_{k \ell m}$ field canonical. We omit writing the contribution to eq.\,(\ref{eq: expansion}) from solutions describing the bound states of the system, as we are interested in the region $E \geq 0$. The solution $\psi_{k \ell m}(\vec{r})$ is analytically obtained as\,\cite{landau}
\begin{align}
    \psi_{k \ell m}(\vec{r}) =
    \frac{\Gamma(1 + \ell + i\alpha m_e/(2k))}{(2\ell + 1)!r}
    \exp\left(\frac{\pi \alpha m_e}{4k}\right)
    M(i\alpha m_e/k, \ell + 1/2, -2ikr)
    Y_{\ell m}(\theta,\varphi),
\end{align}
Here, $Y_{\ell m}(\theta, \varphi)$, $M(a,b,c)$, and $\Gamma(x)$ are the spherical harmonic, the Whittaker (of the first kind), and the Gamma functions. Substituting the expansion\,(\ref{eq: expansion}) into eq.\,(\ref{eq: pNR Lagrangian}) gives
\begin{align}
    \mathcal{L}^{(\text{pNR})}_\mathfrak{f}
    \simeq
    &
    -\frac{1}{2} \vec{\mathfrak{f}}
    \,(\Box + m_\mathfrak{f}^2)\,
    \vec{\mathfrak{f}}
    +\int \frac{dk}{2\pi}\,\vec{C}^\dagger_{k00}
    \left[
        i\partial_0 + \frac{\nabla^2}{4m_e} - \frac{k^2}{m_e}
    \right] \vec{C}_{k00}
    \label{eq: pNR Lagrangian 2}
    \\
    &
    -\sqrt{2} (g_{\rm B-L} - \epsilon e)\,
    \vec{\mathfrak{f}} \cdot
    \int \frac{dk}{2\pi} \left\{ \frac{\alpha m_e k}{1 - \exp\,(-\pi \alpha m_e/k)} \right\}^{1/2}
    \left[
        e^{2 i m_e x^0}\vec{C}^\dagger_{k00}
        +
        e^{-2 i m_e x^0}\vec{C}_{k00}
    \right],
    \nonumber
\end{align}
where we omitted writing the fields with $\ell \neq 0$, as their wave functions vanish at the origin ($\vec{r} = 0$), and they do not couple to the Féeton dark matter directly at the leading order.

With this Lagrangian\,(\ref{eq: pNR Lagrangian 2}), the decay width of the Féeton dark matter into an electron and a positron is obtained using the self-energy of the dark matter field $\Sigma\,(q^2)$ as follows:
\begin{align}
    \Gamma\,(\mathfrak{f} \to e^- e^+) =
    -\frac{1}{m_\mathfrak{f}}{\rm Im}\,
    \left[ \Sigma\,(m_\mathfrak{f}^2) \right],
\end{align}
where $\Sigma\,(q^2)$ is defined as $\int d^4x\,e^{i q x} \braket{0| T[\mathfrak{f}^i(x) \mathfrak{f}^j(0)] |0} \equiv i \delta^{ij}/[q^2 - m_\mathfrak{f}^2 - \Sigma\,(q^2)]$. The imaginary part of the self-energy originates in the propagator of the two-body state $\vec{C}_{k00}(x)$, $[i\partial_0 + \nabla^2/(4m_e) - k^2/m_e + i0^+]^{-1}$. Then, the decay width is analytically obtained as
\begin{align}
    \Gamma(\mathfrak{f} \to e^- e^+) =
    \frac{(g_{\rm B-L} - \epsilon\,e)^2 \alpha\,m_e^2}
    {2 m_\mathfrak{f}
    \left[
        1 - \exp
        \left(
            -\pi \alpha /\sqrt{m_\mathfrak{f}/m_e - 2}
        \right)
    \right]}
    \label{eq: Sommerfeld electron}
\end{align}
When the effect of the Coulomb force vanishes (i.e., $\alpha \to 0$), the decay width becomes $[(g_{\rm B-L} - \epsilon e)^2/\sqrt{2}]\,m_\mathfrak{f} \sqrt{1 - 2m_e/m_\mathfrak{f}}$, coinciding with the width in eq.\,(\ref{eq: decay width electron}) with $m_\mathfrak{f} \to 2m_e$.

\subsection{Reevaluating the constraints}

\begin{figure}[t]
    \begin{center}
    \subfigure[\qquad\quad$|g_{\rm B-L} - \epsilon e|/g_{\rm B-L}=1$]{
    \includegraphics[width=.482\columnwidth]{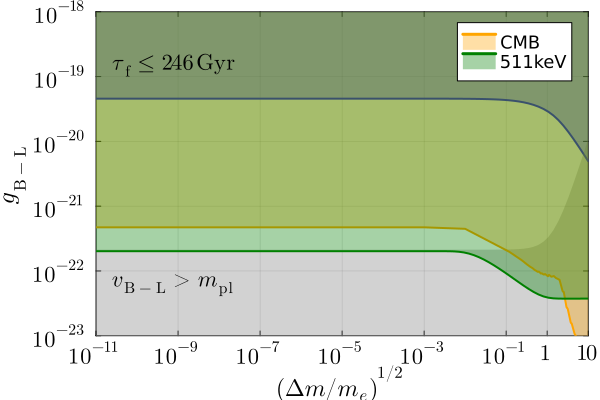}
    }
    \subfigure[\qquad\quad$|g_{\rm B-L} - \epsilon e|/g_{\rm B-L}=10^{-1}$]{
    \includegraphics[width=.482\columnwidth]{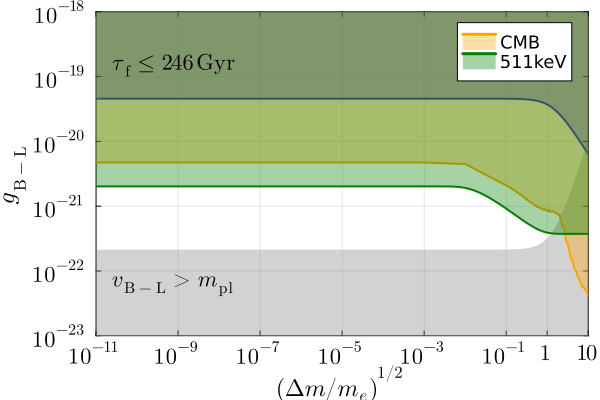}
    }
    \subfigure[\qquad\quad$|g_{\rm B-L} - \epsilon e|/g_{\rm B-L}=10^{-2}$]{
    \includegraphics[width=.482\columnwidth]{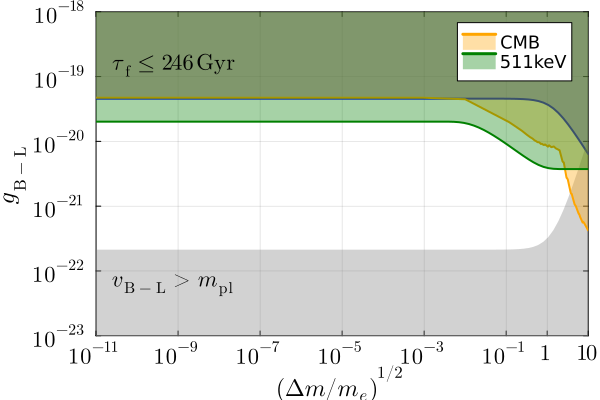}
    }
    \subfigure[\qquad\quad$(\Delta m/m_{e})^{1/2}=10^{-3}$]{
    \includegraphics[width=.482\columnwidth]{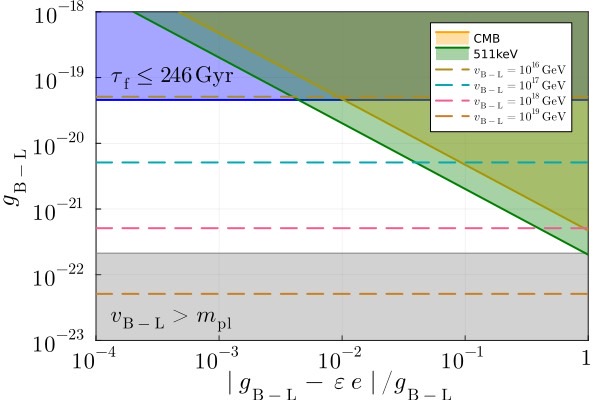}
    }
    \caption{\small \sl The same as those in Fig.\,\ref{fig:result_without_Sommerfeld}, except the decay width of the Féeton dark matter into an electron and a positron\,(\ref{eq: Sommerfeld electron}) is adopted, including the Sommerfeld effect, to depict the region. In addition, in the bottom-right panel, several contours of $v_{\rm B-L}$ are also shown as dashed lines.}
    \label{fig:result_with_Sommerfeld}
    \end{center}
\end{figure}

Let us reevaluate the constraints on the Féeton dark matter parameters discussed in the previous section, using the partial decay width of the dark matter into an electron and a positron with the threshold (Sommerfeld) effect. Among various constraints addressed in section\,\ref{sec: Detection}, those from the CMB and the 511\,keV observations are relevant to the $\mathfrak{f} \to e^- e^+$ decay process. On the other hand, other constraints such as that in eq.\,(\ref{eq: vev}) and those on the decay width into a neutrino pair are not affected by the modification of the $\mathfrak{f} \to e^- e^+$ decay width. Then, the model parameter region consistent with the limits shown in Fig.\,\ref{fig:result_without_Sommerfeld} is modified when we include the Sommerfeld effect on $\Gamma(\mathfrak{f} \to e^- e^+)$, as shown in Fig.\,\ref{fig:result_with_Sommerfeld}.

First, as seen in the top and bottom-left panels of Fig.\,\ref{fig:result_with_Sommerfeld}, the limits from the CMB and 511\,keV observations are independent of $\Delta m/m_e$ in the threshold region, unlike those seen in Fig.\,\ref{fig:result_without_Sommerfeld}. This is because, as shown in eq.\,(\ref{eq: Sommerfeld electron}), $\Gamma(\mathfrak{f} \to e^- e^+) \simeq (g_{\rm B-L} - \epsilon e)^2\,\alpha m_e/4$ at the region, so it is independent of $\Delta m/m_e = (m_\mathfrak{f} - 2m_e)/m_e$, meaning that the decay width of the Féeton dark matter into $e^- e^+$ never vanishes even at the deep threshold region. Next, as seen in the top-left panel of the figure, the model with $\epsilon = 0$ is marginally excluded due to the Sommerfeld effect, meaning it requires a non-zero $\epsilon$ to find a viable parameter region. Third, as seen in the bottom-right panel, $|g_{\rm B-L} - \epsilon e|/g_{\rm B-L} \lesssim 10^{-2}$ is required when  $v_{\rm B-L} = {\cal O}(10^{16})$\,GeV, i.e., the U(1)$_{\rm B-L}$ breaking scale explaining observed tiny neutrino masses naturally (GUT scale seesaw). This discussion holds when the masses of right-handed neutrinos originate in the vacuum expectation value of the Abelian Higgs field $\Phi$ having the B-L charge two, i.e., when active neutrino masses are given by $(y^{(\nu)} v_{\rm EW})^2/(y^{(N)} v_{\rm B-L})$, as deduced from eq.\,(\ref{eq: lagrangian}). On the other hand, when the U(1)$_{\rm B-L}$ breaking is driven by the other Abelian Higgs field $\varphi$ having the B-L charge one, the right-handed neutrino masses are from the higher-dimensional interaction, $y^{(N)}\varphi^2 N^2/m_{\rm pl}$, so the active neutrino masses are given by $(y^{(\nu)} v_{\rm EW})^2\,m_{\rm pl}/(y^{(N)} v_{\rm B-L}^2)$. Then, the natural U(1)$_{\rm B-L}$ breaking scale is modified to $v_{\rm B-L} \sim {\cal O}(10^{-17})$\,GeV, requiring $|g_{\rm B-L} - \epsilon e|/g_{\rm B-L} \lesssim 10^{-1}$. It is worth notifying here that the cancellation between $g_{\rm B-L}$ and $\epsilon e$ is hardly regarded as a fine-tuning among the Lagrangian parameters, for the U(1)$_{\rm B-L}$ model in eq.\,(\ref{eq: lagrangian}) is indistinguishable from the U(1)$_{\rm B-L+xY}$ model, as discussed in appendix\,\ref{app: B-L+xY}: From the viewpoint of the latter model, the above cancellation is simply the matter of the charge assignment associated with the $U(1)_{\rm B-L+xY}$ symmetry.

\section{Summary \& Discussion}
\label{sec: summary and future direction}

We studied the Féeton dark matter predicted in the minimal U(1)$_{\rm B-L}$ model with the kinetic mixing, especially when its mass is above the $e^- e^+$ threshold. After applying all the constraints obtained from astrophysical and cosmological observations as well as theoretical discussions, the viable parameter region is found to be at the threshold region, i.e., $m_\mathfrak{f} \simeq 2m_e$. So, to properly estimate the partial decay width of the dark matter into a pair of electrons, i.e., $\Gamma(\mathfrak{f} \to e^-e^+)$, we have included the Sommerfeld effect in its calculation. Then, we found that the result obtained by including the Sommerfeld effect is very different from that without including the effect: First, the decay of the Féeton dark matter into $e^-e^+$ never vanishes even at the deep threshold region, i.e., $\Delta m/m_e \ll 1$. Next, as a result, the model without the kinetic mixing, i.e., $\epsilon = 0$, is excluded, and a non-zero kinetic mixing is needed to find a viable parameter region, i.e., to suppress the decay width $\Gamma(\mathfrak{f} \to e^-e^+)$ (via the cancellation between the U(1)$_{\rm B-L}$ gauge coupling $g_{\rm B-L}$ and the kinetic mixing parameter $e\epsilon$).\footnote{
    The model with $\epsilon = 0$ still survives when the Féeton dark matter is not comprised of 100\,\% of the dark matter observed today\,\cite{Lin_2022}. In such a case, the dark matter is not necessarily in the threshold region and has "a few MeV" mass. Then, the in-flight annihilation (gamma-ray) signal caused by the propagation of a positron from the decay (before forming Ps) could be the one that discriminates between this and our scenarios.}
Finally, the viable parameter region of the Féeton dark matter is found to be consistent with the natural sale of the U(1)$_{\rm B-L}$ breaking, i.e., the so-called GUT scale seesaw.

It is also interesting to consider how we can test the Féeton dark matter scenario in the future. One possibility is to observe the 21\,cm power spectrum. Red-shifted 21\,cm line signals contain information about the intergalactic medium from the cosmic dawn to the reionization epoch. The inhomogeneous decay of the dark matter injects energy into the medium and influences the power spectrum. Future observations of the spectrum by, e.g., the Hydrogen Epoch of Reionization Array (HERA) would be sensitive to the decay width into $e^-e^+$ at the level of $\Gamma(\mathfrak{f} \to e^- e^+)  < 10^{-28}$\,s$^{-1}$\,\cite{Sun:2023acy}, which is $10^3$ better than the present 511\,keV constraint. The other possibility is updated 511\,keV observations. Future telescopes will reveal the morphology of the 511\,keV emission at the galactic center, which may identify several sources of the positron injection, making the dark matter detection via the 511\,keV observation more sensitive. In particular, observing the 511\,keV emission at the outer edge of the galactic center is interesting. Because the Féeton dark matter signal is expected to be almost proportional to the dark matter profile, which could substantially extend to the edge, the dark matter signal at the edge resembles the isotropic diffused 511\,keV emission. Such emission is difficult to detect by a hard X-ray telescope using a coded aperture mask, while it can be done by a sensitive Compton (MeV gamma-ray) telescope such as the Compton Spectrometer and Imager (COSI)\,\cite{Tomsick:2021wed, Aramaki:2022zpw, Tomsick:2023aue}. We leave this discussion for future study.

\appendix

\section{\texorpdfstring{U(1)$_{\rm B-L}$}{} model and \texorpdfstring{U(1)$_{\rm B - L + xY}$}{} model}
\label{app: B-L+xY}

We confirm that the U(1)$_{\rm B-L}$ extension of the SM associated with the kinetic mixing term between the field strength tensors of the U(1)$_{\rm B-L}$ and the hypercharge gauge bosons, i.e., the model defined in eq.\,(\ref{eq: lagrangian}), is indistinguishable from the U(1)$_{\rm B - L + xY}$ extension of the SM that are not associated with such a kinetic mixing term, when the kinetic mixing term in the U(1)$_{\rm B-L}$ model, $\xi$, is small enough as well as the parameter $x$ is appropriately chosen\,\cite{Lee:2016ief}.

The covariant derivatives acting on the SM fermions, SM Higgs boson, right-handed neutrinos, and new "Higgs" boson $\Psi$, in the U(1)$_{\rm B-L}$ model are in general given by 
\begin{align}
    D_\mu =
    \partial_\mu
    - i
    \left(
        g_S G_\mu
        + g W_\mu
        + g' Y B_\mu
        + g_{\rm B-L} q V_\mu
    \right),
\end{align}
where $g_s$ is the strong coupling, $G_\mu \equiv G_\mu^a (\lambda^a/2)$ with $G_\mu^a$ and $\lambda^a$ being the gluon field and the Gell-Mann matrix, , $W_\mu \equiv W_\mu^a (\tau^a/2)$ with $\tau^a$ being the Pauli matrix, $Y$ is the hypercharge, and $q$ is the B-L charge. Using the field redefinition of the hypercharge gauge boson,
\begin{align}
    B_{\mu} \rightarrow
    B_{\mu} + \frac{g_{\rm B-L}}{g'}xV_\mu,
\end{align}
the above covariant derivative is written as that of the U(1)$_{\rm B-L+xY}$ extension of the SM,
\begin{align}
    D_\mu \rightarrow D_\mu =
    \partial_\mu
    - i
    \left[
        g_S G_\mu
        + g W_\mu
        + g' Y B_\mu
        + g_{\rm B-L} (q + x Y) V_\mu
    \right].
\end{align}
This field redefinition changes the kinetic terms of the relevant gauge fields as follows:
\begin{align}
    \mathcal{L}_{\rm B-L} &\supset
    -\frac{1}{4}V_{\mu\nu} V^{\mu\nu}
    -\frac{1}{4}B_{\mu\nu} B^{\mu\nu}
    -\frac{\xi}{2} B_{\mu\nu} V^{\mu\nu}
    \nonumber \\
    &\rightarrow
    -\frac{1}{4}
    \left(
        1 
        + 2 \frac{g_{\rm B-L}}{g'} x \xi
        + \frac{g^2_{\rm B-L}}{g'^2} x^2
    \right) V_{\mu\nu} V^{\mu\nu}
    -
    \frac{1}{2}
    \left(
        \frac{g_{\rm B-L}}{g'} x
        +\xi
    \right) B_{\mu\nu} V^{\mu\nu}
    -\frac{1}{4}B_{\mu\nu}B^{\mu\nu}.
\end{align}
So, taking $x = - (g'/g_{\rm B-L})\,\xi$ annihilates the kinetic mixing between $B_{\mu\nu}$ and $V_{\mu\nu}$, while the kinetic terms of the relevant gauge fields are intact up to ${\cal O}(\xi)$, as $g_{\rm B-L}\,x = {\cal O}(\xi)$.

\section{Deriving the pNR Lagrangian}
\label{app: about pNR}

When considering $e^\pm$ around the threshold region, the momentum transfer among $e^\pm$ carried by exchanged photons is soft, and the propagating electrons are almost on-shell. Integrating out all the fields except those of the Féeton dark matter and the potential part of the electron $\psi_{\rm Pot}(x)$ from the Lagrangian\,(\ref{eq: lagrangian}),\footnote{
    Since we are interested in the leading order contribution to the decay width $\Gamma(\mathfrak{f} \to e^- e^+)$, we also integrate the ultra-soft part of the electron field and all the parts of the photon field out from the Lagrangian\,(\ref{eq: lagrangian}). } 
and expanding $\psi_{\rm Pot}(x)$ in terms of its velocity as
\begin{align}
    \psi_{\rm Pot}(x) =
    \begin{pmatrix}
        e^{- i m_e x^0} \eta(x) + i e^{i m_e x^0} [\vec{\nabla} \cdot \vec{\sigma}\,\xi(x)]/(2 m_e) + \cdots \\
        e^{i m_e x^0} \xi(x) - i e^{-i m_e x^0} [\vec{\nabla} \cdot \vec{\sigma}\,\eta(x)]/(2 m_e) + \cdots
    \end{pmatrix},
    \label{eq: NR expansion}
\end{align}
with $\sigma$ being the Pauli matrix, gives us the so-called non-relativistic Lagrangian as follows:
\begin{align}
    \mathcal{L}^{({\rm NR})}_\mathfrak{f}
    &\simeq
    \frac{1}{2} \mathfrak{f}_\mu \left[ (\Box + m_\mathfrak{f}^2)\,g^{\mu\nu} - \partial^\mu \partial^\nu \right] \mathfrak{f}_\nu
    +
    \eta^\dagger \left( i\partial_{x^0} + \frac{\nabla^2}{2m_e} \right) \eta
    +
    \xi^\dagger \left( i\partial_{x^0} - \frac{\nabla^2}{2m_e} \right) \xi
    \nonumber \\
    &+
    \frac{1}{2} \int d^4y \frac{\alpha\,\delta(x^0 - y^0)}{|\vec{x}-\vec{y}|}
    \left[ \eta^\dagger(x)\,\vec{\sigma}\,\xi(y) \right]
    \cdot
    \left[ \xi^\dagger(y)\,\vec{\sigma}\,\eta(x) \right]
    \nonumber \\
    &-
    (g_{\rm B-L} - \epsilon e)\,\vec{\mathfrak{f}}
    \cdot
    [e^{2 i m_e x^0} \eta^\dagger\,\vec{\sigma}\,\xi
    +
    e^{-2 i m_e x^0} \xi^\dagger \vec{\sigma} \eta].
    \label{eq: NR Lagrangian}
\end{align}
$\eta\,(\eta^\dagger)$ is the operator annihilating\,(creating) a non-relativistic electron $e^-$, while $\xi\,(\xi^\dagger)$ creates\,(annihilates) its anti-particle $e^+$. In the above Lagrangian, we drop several terms describing other states not directly coupled to the Féeton dark matter, such as $\eta^\dagger(x) \xi(y)$. After introducing the two-body fields describing the $e^- e^+$ two-body system that directly couples to the Féeton dark matter field $\mathfrak{f}$, and integrating the component fields $\eta$ and $\xi$ in the above Lagrangian\,(\ref{eq: NR Lagrangian}), we have the potential NR Lagrangian for the two-body system:
\begin{align}
    \mathcal{L}^{({\rm pNR})}_\mathfrak{f}
    &\simeq 
    -\frac{1}{2} \vec{\mathfrak{f}}\,
    (\Box + m_\mathfrak{f}^2)\,\vec{\mathfrak{f}}
    +\int d^3r\,\vec{\Phi}^\dagger(\vec{r},x) \left(
        i\partial_{x^0}
        +\frac{\nabla_x^2}{4m_e}
        +\frac{\nabla_r^2}{m_e}
        +\frac{\alpha}{r}
    \right) \vec{\Phi}(\vec{r},x)
    \nonumber \\
    &
    -\sqrt{2} (g_{\rm B-L} - \epsilon e)\,\vec{\mathfrak{f}}
    \cdot
    \left[
        e^{2i m_e x^0} \vec{\Phi}^\dagger(\vec{0}, x)
        +
        e^{-2i m_\mu x^0} \vec{\Phi}(\vec{0},x)
    \right],
\end{align}
where $\vec{\mathfrak{f}}$ is the spatial components of the Féeton dark matter field, while $\vec{\Phi}(\vec{r},x)$ and $\vec{\Phi}^\dagger(\vec{r},x)$ are the fields annihilating and creating the $e^- e^+$ two-body system with the total spin of one (the spin-triplet state), respectively. Here, $\vec{r}$ and $x$ are the relative and center-of-mass coordinates of the two-body system. See also Ref.\,\cite{Matsumoto:2022ojl} for a more detailed derivation.



\section*{Acknowledgments}

We want to thank and acknowledge Thomas Siegert (U. Würzburg) for pointing out a potential difference between the annihilation and production rates of the positronium. Without his suggestion, it would be impossible to put a robust constraint on the dark matter from the 511keV observation. This work is supported by Grant-in-Aid for Scientific Research from the Ministry of Education, Culture, Sports, Science, and Technology (MEXT), Japan; 23K20232\,(20H01895), 20H00153, 24H00244 (for S. Matsumoto), 24H02244 (for S. Matsumoto and T. T. Yanagida), by JSPS Core-to-Core Program; JPJSCCA20200002 (for S. Matsumoto), by the China Grant for Talent Scientific Start-Up Project and by the Natural Science Foundation of China (NSFC); 12175134 (for T. T. Yanagida), and by World Premier International Research Center Initiative (WPI), MEXT, Japan (Kavli IPMU).

\bibliographystyle{unsrt}
\bibliography{references}

\end{document}